\documentclass[conference]{IEEEtran}
\usepackage{blindtext}
\usepackage{graphicx}
\usepackage{bigints}
\usepackage{caption}
\newlength\figureheight
\newlength\figurewidth
\usepackage{cite}
\usepackage{url}
\usepackage{blindtext, graphicx}
\usepackage{amsmath,amssymb,mathtools,bm,etoolbox}
\newcommand\Mark[1]{\textsuperscript#1}
\usepackage{cuted}
\usepackage{multicol,lipsum}
\usepackage{multirow}
\usepackage{breqn}

\newcommand{\e}[1]{{\mathbb E}\left[ #1 \right]}
\usepackage{stackengine}
\def\delequal{\mathrel{\ensurestackMath{\stackon[1pt]{=}{\scriptstyle\Delta}}}}
\usepackage{suffix}

\DeclarePairedDelimiterX\MeijerM[3]{\lparen}{\rparen}%
{\begin{smallmatrix}#1 \\ #2\end{smallmatrix}\delimsize\vert\,#3}

\newcommand\MeijerG[8][]{%
  G^{\,#2,#3}_{#4,#5}\MeijerM[#1]{#6}{#7}{#8}}

\WithSuffix\newcommand\MeijerG*[7]{%
  G^{\,#1,#2}_{#3,#4}\MeijerM*{#5}{#6}{#7}}
\begin{document}

\title{Mixed RF/FSO Relaying Systems with Hardware Impairments}
\author{Elyes~Balti\Mark{1},~Mohsen~Guizani\Mark{1},~Bechir~Hamdaoui\Mark{2} and Bassem Khalfi\Mark{2}\\
        \Mark{1}University of Idaho, USA,  \Mark{2}Oregon State University, USA}

\maketitle

\begin{abstract}
In this work, we provide a detailed analysis of a dual-hop fixed gain (FG) amplify-and-forward relaying system, consisting of a hybrid radio frequency (RF) and free-space optical (FSO) channels. We introduce an impairment model which is the soft envelope limiter (SEL). Additionally, we propose the partial relay selection (PRS) protocol with outdated channel state information (CSI) based on the knowledge of the RF channels in order to select one relay for the communication. Moreover, the RF channels of the first hop experience Rayleigh fading while we propose a unified fading model for the FSO channels, called the unified Gamma Gamma (GG), taking into account the atmospheric turbulence, the path loss and the misalignment between the transmitter and the receiver aperture also called the pointing error. Novel closed-forms of the outage probability (OP), the bit error probability (BEP) and the average ergodic capacity (EC) are derived in terms of Meijer-G and Fox-H functions. Capitalizing on these metrics, we also derive the asymptotical high signal-to-noise ratio (SNR) in order to get engineering insights into the impacts of the hardware impairments and the system parameters as well. Finally, using Monte Carlo simulations, we validate numerically the derived mathematical formulations.
\end{abstract}
\begin{IEEEkeywords}
Partial relay selection, outdated CSI, amplify-and-forward, pointing error, path loss, hardware impairments.
\end{IEEEkeywords}

\IEEEpeerreviewmaketitle

\section{Introduction}
With the rapid increase of Internet mobile stations and the high demands for bandwidth, RF cellular networks have reached a bottleneck due to the limited access to spectrum resources. In addition, existing backhaul network infrastructure, which connects the core network to the edges, cannot support the massive flows of data traffic. Recently, researchers have proposed to use optical fibers (OF) as a solution for alleviating the backhaul load congestion. However, as the number of cells becomes very large (e.g., in ultra dense cellular networks), networks can still suffer from the limited use of the OF, where cable installations are very costly and sometimes even restricted \cite{1}. To support the large number of users, FSO (free-space optical) technology emerges as an alternative or complementary solution to the RF and OF links, since it is more flexible, license-free, power efficient, cost effective, no installation restriction and most importantly it increases the capacity of cellular networks \cite{2}. Because of these advantages, FSO became a timely research topic and a promising technique, which has recently gained enormous interest especially for the mixed RF/FSO systems. However, there is an inevitable limitation of such systems, namely the low signal coverage that some cellular areas may experience. In order to increase the scalability and the coverage of the cellular networks, relays can be implemented as intermediate nodes between the source $S$ and the destination $D$ to assist and amplify the signal over long distances. Additionally, relays can be installed in any cells/areas and with large numbers since they are of low cost.\\
Relaying communications can be classified into multiple categories. The most widespread schemes are amplify-and-forward (AF) \cite{10,10-1,12} and decode-and-forward (DF) \cite{5}. In case of AF or non-regenerative relaying, the relay only amplifies the received signal by either fixed or variable gain and then forwards it whereas the DF or regenerative relaying decodes the received signal to reduce the noise, re-encode it and then retransmits the signal again. Unlike the vast majority of work in the literature, in practice, the relays are not ideal hardware and are susceptible to the imperfections during the signal amplification due to its low quality. In fact, these imperfections are eventually created by the non-linear high power amplification \cite{6} and are classified into many types, depending on the amplifier nature \cite{7}. It has been shown previously in \cite{8} that the High Power Amplifier (HPA) non-linearities create non-linear distortion which creates not only irreducible floors for the outage probability and the bit error probability but also saturates the system capacity by a destructive ceiling. Although it is very challenging to consider non-ideal hardware, the work mentioned here considered impaired hardware over full RF systems with single relay \cite{9}.\\
Our contribution is to quantify the impacts of the HPA non-linearities on the mixed RF/FSO system with multiple relays. Although Balti \textit{et. al} \cite{10,10-1} introduced the hardware impairments to the mixed RF/FSO system, they considered a general model of hardware impairments. In this work, we introduce a more specific impairment model which is the soft envelope limiter (SEL) HPA non-linearities to the system. Additionally, we assume the intensity modulation and direct detection (IM/DD) for signal reception and since the RF channels are time-varying, the partial relay selection (PRS) with outdated channel state information (CSI) is adopted to select one relay for the signal forwarding. Furthermore, we suggest the unified Gamma Gamma (GG) for the FSO channels model considering the pointing error, the path loss attenuation and the atmospheric turbulence related to the weather state (clear, light/moderate fog, moderate/heavy rain and hazy). The rest of this paper is structured as follows: Section II describes the system and channel models while the outage probability, the bit error probability and the ergodic capacity analysis are given in Section III. Numerical and analytical results are discussed in Section IV. Finally, the concluding remarks are reported in Section V.
\section{System and Channel Models}
\subsection{System Model}
In this system, $S$ communicates with $D$ by selecting one relay among an intermediate set of $N$ relays. To achieve this selection, the PRS protocol states that $S$ receives the CSIs from the relays, sorts them in an increasing order and then selects the relay/channel with the highest CSI. Given that the RF channels are time-varying, the incoming CSIs instantaneously fluctuate before and after the relay selection. In this case, the highest SNR link before the selection does not remain the same after the selection and thereby the selection will be achieved based on outdated CSI. Moreover, since the relays operate at the half-dupplex mode, the best relay may not be always available for the transmission. In this case, $S$ will select the next best available relay and so on so forth. To model the relation between the outdated and the updated CSIs, we associate the correlation coefficient $\rho$ as follows:
\begin{equation} \label{eq:1}
h_{1(m)} = \sqrt{\rho}~\hat{h}_{1(m)} + \sqrt{1-\rho}~w_{m}
\end{equation}
where $w_{m}$ is a random variable that follows the circularly complex Gaussian distribution with the same variance of the channel gain $h_{1(m)}$. The correlation coefficient $\rho$ is given by the Jakes' autocorrelation model as follows:
\begin{equation}
\rho = J_0(2\pi f_{d} T_d)
\end{equation}
where $J_0(\cdot)$ is the zeroth order Bessel function of the first kind, $T_d$ is the time delay between the current and the delayed CSI versions and $f_d$ is the maximum Doppler frequency of the channels.\\
Supposing that $S$ selects the relay with rank $m$, the amplification gain can be given by:
\begin{equation}
G = \sqrt{\frac{\sigma_p^2}{\e{|h_m|^2} P_s + \sigma^2_1}}
\end{equation}
where $\e{\cdot}$ is the expectation operator, $P_s$ is the average transmitted power from $S$, $\sigma^2_1$ is the noise power and $\sigma_p^2$ is the mean power of the signal at the output of the relay block. For a given saturation level $A_{sat}$, the amplifier operates at a certain input back-off (IBO), which is given by:
\begin{equation}
\text{IBO} = \frac{A^2_{sat}}{\sigma_p^2}
\end{equation}
Since the HPA creates a non-linear distortion, we refer to the Bussgang linearization theory to linearize the distortion \cite{busg}. Thus, the output of the nonlinear circuit can be given by \cite[Eq.~(9)]{12}:
\begin{equation}
\Omega_m = \nu~x + b
\end{equation}
where $\nu$ is the scale of the input signal and $b$ is an uncorrelated non-linear distortion with the input signal that follows the Gaussian distribution $b \backsim \mathcal{CN} (0,~\sigma^2_b)$. For the SEL model, the parameters $\nu$, $\sigma^2_b$ and the clipping factor $\mu$ are given by:
\begin{equation}\label{eq:17}
\begin{split}
\nu = 1 - \exp\left(-\frac{A^2_{sat}}{\sigma_p^2}\right) + \frac{\sqrt{\pi}A_{sat}}{2\sigma_p}\text{efrc}\left(-\frac{A_{sat}}{\sigma_p}\right)\\
\sigma^2_{b} = \sigma_p^2\left[1 - \exp\left(-\frac{A^2_{sat}}{\sigma_p^2}\right)-\nu^2\right]~~~~~~~~~~~~~~~~~~\\
\mu = 1 - \exp\left(-\frac{A^2_{sat}}{\sigma_p^2}\right)~~~~~~~~~~~~~~~~~~~~~~~~~~~~~~~
\end{split}
\end{equation}
where \text{erfc}$(\cdot)$ is the complementary error function.\\
The average transmitted power at the relay can be given in terms of the clipping factor as follows:
\begin{equation}
P_t = \mu \sigma_p^2
\end{equation}
The relays employ the subcarrier intensity modulation (SIM) for the electrical to optical conversion.\\
The overall signal-to-noise plus distortion ratio (SNDR) is given by:
\begin{equation}
\gamma_{ni} = \frac{\gamma_{1(m)}\gamma_{2(m)}}{\kappa\gamma_{2(m)} + \e{\gamma_{1(m)}} + \kappa}
\end{equation}
where $\kappa$ is defined by:
\begin{equation}
\kappa = 1 + \frac{\sigma^2_{b}}{\nu^2G^2\sigma_1^2}
\end{equation}
Note that for the case of ideal relays ($\kappa$ = 1), Eq.~(8) is reduced to:
\begin{equation}
\gamma_{id} = \frac{\gamma_{1(m)}\gamma_{2(m)}}{\gamma_{2(m)} + \e{\gamma_{1(m)}} + 1}
\end{equation}
which is the expression of the overall SNR of an ideal relaying system.\\
The instantaneous SNR of the first hop can be given by:
\begin{equation}
\gamma_{1(m)} = \frac{|h_{m}|^2P_s}{\sigma^2_1}
\end{equation}
The average SNR of the first hop can be expressed as follows:
\begin{equation}
\overline{\gamma}_1 = \frac{P_s}{\sigma^2_1}
\end{equation}
The instantaneous SNR of the $m$-th optical channel can be obtained by:
\begin{equation}
\gamma_{2(m)} = \frac{|I_m|^2\eta^2P_t^2}{\sigma_2^2}
\end{equation}
where $I_m$, $\eta$ and $\sigma^2_2$ are the $m$-th channel gain, the optical-electrical conversion and the noise power of the optical channel, respectively.
\subsection{Channels Model}
Since the RF channels of the first hop are subject to the Rayleigh fading, the instantaneous SNRs $\gamma_{1(m)}$ and $\hat{\gamma}_{1(m)}$ are two jointly exponential random variables. The joint probability density function (PDF) of $\gamma_{1(m)}$ and $\hat{\gamma}_{1(m)}$ is expressed as follows:
\begin{equation}
f_{\gamma_{1(m)},\hat{\gamma}_{1(m)}}(x,y) = \frac{1}{(1-\rho)\overline{\gamma}^2_1}e^{-\frac{x+y}{(1-\rho)\overline{\gamma}_1}}I_0\left(\frac{2\sqrt{\rho xy}}{(1-\rho)\overline{\gamma}_1}  \right)
\end{equation}
where $I_\nu(\cdot)$ is the $\nu$-th order modified Bessel function of the first kind.\\
After some mathematical manipulations, the cumulative distribution function (CDF) of $\gamma_{1(m)}$ is given by:
\begin{equation}
\begin{split}
F_{\gamma_{1(m)}}(x) =&~ 1 - m{N \choose m}\sum_{n=0}^{m-1} \frac{(-1)^n}{N-m+n+1}{m-1 \choose n}  \\&
\times~\exp\left(-\frac{ (N-m+n+1)x }{[(N-m+n)(1-\rho)+1]\overline{\gamma}_1}\right)~
\end{split}
\end{equation}
The expectation $\e{\gamma_{1(m)}}$ can be expressed as follows:
\begin{equation}
\begin{split}
\e{\gamma_{1(m)}} =&~ m{N \choose m}\sum_{n=0}^{m-1} {m-1 \choose n} (-1)^n\\& \times~\frac{[(N-m+n)(1-\rho)+1]\overline{\gamma}_1}{(N-m+n+1)^2}
\end{split}
\end{equation}
Regarding the channels of the second hop, the channel gain $I_m$ can be expressed as follows:
\begin{equation}
I_m = I_a I_l I_p
\end{equation}
where $I_a, I_l$ and $I_p$ are the atmospheric tubulence, the path loss and the pointing error, respectively. The table below summarizes the main parameters of the optical channel.
\begin{table}[h!]
\renewcommand{\arraystretch}{1.3}
\caption{Parameters of the optical channel}
\label{tab:example}
\centering
\begin{tabular}{|c|c|}
    \hline
    Parameter  &  Symbol\\
    \hline
    \hline
    Weather attenuation    &   $\sigma$\\
    \hline
    Jitter variance    & $\sigma^2_s$   \\
    \hline
    Rytov variance & $\sigma^2_R$\\
    \hline
    Wave number    & $k$   \\
    \hline
    Wavelength    & $\lambda$   \\
    \hline
    Pointing error coefficient    & $\xi$   \\
    \hline
    Beam waist at the relay & $w_0$\\
    \hline
    Beam waist    & $w_{L}$   \\
    \hline
    Equivalent beam waist    & $w_{Leq}$   \\
    \hline
    Length of the optical link    & $L$   \\
    \hline
    Radius of the receiver aperture    & $a$   \\
    \hline
    Fraction of the collected power at $L = 0$    & $A_0$   \\
    \hline
    Radius of curvature & $F_0$\\
    \hline
    Refractive index of the medium & $C_n^2$\\
    \hline
    Small scale turbulence & $\alpha$\\
    \hline
    Large scale turbulence & $\beta$\\
    \hline
    Radial displacement of the beam at the receiver & $R$\\
    \hline
\end{tabular}
\end{table}
\\
Using the Beers-Lambert law, the path loss can be expressed as follows:
\begin{equation}
I_l = \exp(-\sigma L)
\end{equation}
The pointing error $I_p$ made by Jitter can be given as:
\begin{equation}
I_p = A_0 \exp\left(-\frac{2R^2}{w^2_{Leq}}  \right)
\end{equation}
Assuming that the radial displacement of the beam at the detector follows the Rayleigh distribution, the PDF of the pointing error can be expressed as follows:
\begin{equation}
f_{I_p}(I_p) = \frac{\xi^2}{A_0^{\xi^2}}I^{\xi^2-1}_p~,~~0\leq I_p\leq A_0
\end{equation}
The small and large atmospheric scales can be determined by:
\begin{equation}
\begin{split}
\alpha = \left( \exp\left[ \frac{0.49 \sigma_R^2}{(1+1.11\sigma_R^{\frac{12}{5}})^{\frac{7}{6}}}\right] -1\right)^{-1}\\
\beta = \left( \exp\left[ \frac{0.51 \sigma_R^2}{(1+0.69\sigma_R^{\frac{12}{5}})^{\frac{5}{6}}}\right] -1\right)^{-1}
\end{split}
\end{equation}
where the Rytov variance is given by:
\begin{equation}
\sigma^2_R = 1.23~C^2_nk^{7/6}L^{11/6}
\end{equation}
The pointing error coefficient can be expressed in terms of the Jitter standard deviation and the equivalent beam waist as follows:
\begin{equation}
\xi = \frac{w_{Leq}}{2\sigma_s}
\end{equation}
We can also relate $w_{Leq}$ with the beam width $w_L$ of the Gaussian laser beam at the distance $L$ as follows:
\begin{equation}
w^2_{Leq} = \frac{w^2_L\sqrt{\pi}\text{erf}(v)}{2v\exp(-v^2)},~v =\frac{\sqrt{\pi}a}{\sqrt{2}w_L}
\end{equation}
where \text{erf}$(\cdot)$ is the error function. The fraction of the collected power at the relay $A_0$ is given by
\begin{equation}
A_0 = |\text{erf}(v)|^2
\end{equation}
The Gaussian beam waist itself can be defined as:
\begin{equation}
\begin{split}
w_L = w_0\sqrt{(\Theta_0 + \Lambda_0)(1 + 1.63~\sigma_R^{12/5}\Lambda_1)}~~~\\
\Theta_0 = 1 - \frac{L}{F_0},~~\Lambda_0 = \frac{2L}{kw_0^2},~~\Lambda_1 = \frac{\Lambda_0}{\Theta_0^2 + \Lambda^2_0}
\end{split}
\end{equation}
After some mathematical manipulations, the average electrical SNR of the optical channel can be obtained by:
\begin{equation}
\overline{\gamma}_2 = \frac{P_t^2\eta^2}{\sigma^2_2}h^2A_0^2I_l^2,~~h = \frac{\xi^2}{\xi^2 + 1}
\end{equation}
Since the atmospheric turbulence $I_a$ is modeled as GG, the PDF of the optical irradiance of the $m$-th channel can be given by \cite[Eq.~(13)]{11}:
\begin{equation}
f_{I_m}(I_m) = \frac{\xi^2 \alpha \beta}{A_0 I_l \Gamma(\alpha)\Gamma(\beta)}
G_{1,3}^{3,0} \Bigg(\frac{\alpha \beta I_m}{A_0I_l}~\bigg|~\begin{matrix} \xi^2 \\ \xi^2-1, \alpha - 1, \beta - 1 \end{matrix} \Bigg)
\end{equation}
After some algebraic transformations, the PDF of the instantaneous optical SNR can be derived as follows:
\begin{equation}
f_{\gamma_{2(m)}}(x) = \frac{\xi^2}{2\Gamma(\alpha)\Gamma(\beta)x}G_{1,3}^{3,0} \Bigg(\alpha\beta h \sqrt{\frac{x}{\overline{\gamma}_2}}~\bigg|~\begin{matrix} \xi^2 \\ \xi^2, \alpha, \beta \end{matrix} \Bigg)
\end{equation}
where $G_{p,q}^{m,n}(\cdot)$ is the Meijer's G-function.
\section{Performance Analysis}
\subsection{Outage Probability Analysis}
The outage probability (OP) is interpreted as the probability that the end-to-end SNDR $\gamma_{\text{ni}}$ falls below a certain outage threshold $\gamma_{\text{th}}$. It can be defined as:
\begin{equation}
P_{\text{out}}(\gamma_{\text{th}}) \delequal \text{Pr}[\gamma_{\text{ni}} < \gamma_{\text{th}}]
\end{equation}
where \text{Pr}$[\cdot]$ is the probability measure notation.\\
Substituting the expression of the overall SNDR (8) in Eq.~(30) and after some mathematical manipulations, the OP can be derived as follows:
\begin{equation}
\begin{split}
P_{\text{out}}(\gamma_{\text{th}}) =& 1 - \frac{2^{\alpha+\beta-3}\xi^2m}{\pi \Gamma(\alpha) \Gamma(\beta)}{N \choose m} \sum_{n = 0}^{m-1}\frac{(-1)^n{m-1 \choose n}}{N-m+n+1}\\& \times~
\exp\left(-\frac{(N-m+n+1)\kappa \gamma_{\text{th}}}{((N-m+n)(1-\rho)+1)\overline{\gamma}_{1}}\right)\\& \times~
G_{2,7}^{7,0} \Bigg(\frac{(\alpha \beta h)^2(N-m+n+1)\gamma_{\text{th}}\zeta}{16( (N-m+n)(1-\rho)+1)\overline{\gamma}_1\overline{\gamma}_2}    ~\bigg|~\begin{matrix} \kappa_1 \\ \kappa_2 \end{matrix} \Bigg)
\end{split}
\end{equation}
where $\zeta,~\kappa_1$ and $\kappa_2$ are given by:
\begin{equation}
\begin{split}
\zeta = \e{\gamma_{1(m)}} + \kappa,~~\kappa_1 = \Delta(2: \xi^2+1)~~~~~~~~~~~~~~~~~~~~~~~~~~\\
\kappa_2 = \Delta(2: \xi^2+1),~\Delta(2:\alpha),~\Delta(2:\beta),~0~~~~~~~~~~~~~~~~~~~~~~~\\
\Delta(j:x) \delequal x/j,...,(x+j-1)/j~~~~~~~~~~~~~~~~~~~~~~~~~~~~~~~~~~~~
\end{split}
\end{equation}
To get further insight into the behavior of the outage system at high SNR, we derive a simpler form of an asymptotic expression using the expansion of the Meijer's G-function as follows:
\begin{equation}
\begin{split}
P^{\infty}_{\text{out}}(\gamma_{\text{th}}) \underset{\overline{\gamma}_2 \gg 1}\cong &~ 1 - \frac{2^{\alpha+\beta-3}\xi^2m}{\pi \Gamma(\alpha) \Gamma(\beta)}{N \choose m} \sum_{n = 0}^{m-1}\frac{(-1)^n{m-1 \choose n}}{N-m+n+1}\\&
\times~\exp\left(-\frac{(N-m+n+1)\kappa \gamma_{\text{th}}}{((N-m+n)(1-\rho)+1)\overline{\gamma}_{1}}\right)\\& \times~ \sum_{r = 1}^7 \frac{\prod_{j=1, j \neq r}^7 \Gamma(\kappa_{2,j} - \kappa_{2,r})}{\prod_{l=1}^2 \Gamma(\kappa_{1,l} - \kappa_{2,r})}\\&\times~
\left(\frac{(\alpha \beta h)^2(N-m+n+1)\gamma_{\text{th}}\zeta}{16((N-m+n)(1-\rho)+1)\overline{\gamma}_1\overline{\gamma}_2}\right)^{\kappa_{2,r}}
\end{split}
\end{equation}
\subsection{Bit Error Probability Analysis}
For most binary modulations, BEP can be defined as:
\begin{equation}
\overline{P_e} = \frac{\delta^{\tau}}{2\Gamma(\tau)}\int\limits_0^\infty \gamma^{\tau - 1}e^{-\delta \gamma} F_{\gamma}(\gamma)d\gamma
\end{equation}
Replacing the expression of the OP (31) in Eq.~(34) and after some algebraic manipulations, BEP can be written as:
\begin{equation}
\begin{split}
\overline{P_e} =&~ \frac{1}{2} - \frac{2^{\alpha+\beta-4}m\xi^2}{\pi \Gamma(\alpha) \Gamma(\beta) \Gamma(\tau)}{N \choose m}\sum_{n=0}^{m-1}\frac{(-1)^n{m-1 \choose n}}{N-m+n+1}\\&\times~ \left( \frac{\delta}{\varrho \kappa + \delta}  \right)^{\tau}G_{3,7}^{7,1} \Bigg(\frac{(\alpha \beta h)^2\varrho \zeta}{16(\varrho \kappa + \delta)\overline{\gamma}_2}~\bigg|~\begin{matrix} 1 - \tau,~\kappa_1 \\ \kappa_2 \end{matrix} \Bigg)\\
\end{split}
\end{equation}
Note that $\varrho$ is given by:
\begin{equation}
\varrho = \frac{(N-m+n+1)}{( (N-m+n)(1-\rho)+1)\overline{\gamma}_1}
\end{equation}
The high SNR asymptotic expression of the BEP can be derived using the expansion of the Meijer's G-function as follows:
\begin{equation}
\begin{split}
\overline{P_e}^{\infty} \underset{\overline{\gamma}_2 \gg 1}\cong &~ \frac{1}{2} - \frac{2^{\alpha+\beta-4}m\xi^2}{\pi \Gamma(\alpha) \Gamma(\beta) \Gamma(\tau)}{N \choose m}\sum_{n=0}^{m-1}\frac{(-1)^n{m-1 \choose n}}{N-m+n+1}\\&\times~\sum_{r = 1}^7 \frac{\prod_{j=1, j \neq r}^7 \Gamma(\kappa_{2,j} - \kappa_{2,r})~\Gamma(\tau + \kappa_{2,r})}{\prod_{l=1}^2 \Gamma(\kappa_{1,l} - \kappa_{2,r})}\\&\times~\left( \frac{\delta}{\varrho \kappa + \delta}  \right)^{\tau}
\left(\frac{(\alpha \beta h)^2\varrho \zeta}{16(\varrho \kappa + \delta)\overline{\gamma}_2}\right)^{\kappa_{2,r}}
\end{split}
\end{equation}
\subsection{Ergodic Capacity Analysis}
The average ergodic capacity, expressed in bps/Hz, can be defined as the maximum error-free data rate transmitted by the overall system channels. Considering IM/DD detection, the ergodic capacity can be obtained by:
\begin{equation}
\overline{C} = \e{\log_2\left(1+\frac{e\gamma}{2\pi}\right)}
\end{equation}
Using the integration by part, the ergodic capacity can be formulated as follows:
\begin{equation}
\overline{C} = \frac{e}{2\pi\log(2)}\int\limits_0^\infty \frac{\overline{F_{\gamma}}(\gamma)}{1+\frac{\gamma e}{2\pi}}d\gamma
\end{equation}
where $\overline{F_{\gamma}}(\cdot)$ is the complementary CDF (CCDF).\\
Substituting the CCDF of (31) in Eq.~(39) and after some mathematical manipulations, the ergodic capacity can be derived as follows:
\begin{equation}
\begin{split}
\overline{C} &= \frac{2^{\alpha+\beta-4}\xi^2me}{\pi^2\Gamma(\alpha)\Gamma(\beta)\log(2)}{N \choose m} \sum_{n=0}^{m-1}\frac{(-1)^n{m-1 \choose n}}{(N-m+n+1)^2}\\&\times \frac{( (N-m+n)(1-\rho)+1 )\overline{\gamma}_1}{\kappa}\\&\times
H_{1,0:1,1:2,7}^{0,1:1,1:7,0} \Bigg( \begin{matrix} (0;1,1) \\ - \end{matrix} ~\bigg|~\begin{matrix} (0 , 1) \\ (0 , 1) \end{matrix} ~\bigg|~\begin{matrix} ( \kappa_1 , [1]_7) \\ (\kappa_2 , [1]_7) \end{matrix}~\bigg|~\frac{e}{2\pi\varrho \kappa},~\frac{e \omega}{2\pi\varrho \kappa} \Bigg)
\end{split}
\end{equation}
where $H_{p_1,q_1:p_2,q_2:p_3,q_3}^{m_1,n_1:m_2,n_2:m_3:n_3}(-|(\cdot,\cdot))$ is the bivariate Fox H-function and $[x]_j$ is the vector containing $j$ elements equal to $x$.\\
Since the closed-form formula provides limited engineering insights, we derive a high SNR expression of the ergodic capacity. Substituting the CCDF of (33) in Eq.~(39), the asymptotic channel capacity can be expressed as follows:
\begin{equation}
\begin{split}
\overline{C}^{\infty} \underset{\overline{\gamma}_2 \gg 1}\cong& \frac{2^{\alpha+\beta-3}\xi^2m}{\pi\Gamma(\alpha)\Gamma(\beta)\log(2)} \sum_{n=0}^{m-1}\frac{(-1)^n{m-1 \choose n} }{N-m+n+1}\\& \times~\exp\left(\frac{2\pi \varrho \kappa}{e} \right)\sum_{r = 1}^7 \frac{\prod_{j=1, j \neq r}^7 \Gamma(\kappa_{2,j} - \kappa_{2,r})}{\prod_{l=1}^2 \Gamma(\kappa_{1,l} - \kappa_{2,r})} \\&\times~\Gamma(1+\kappa_{2,r})\Gamma\left(-\kappa_{2,r},\frac{2\pi \varrho \kappa}{e}  \right) \left(\frac{2\pi \omega}{e} \right)^{\kappa_{2,r}}
\end{split}
\end{equation}
where $\Gamma(\cdot,\cdot)$ is the incomplete upper gamma function and $\omega$ is given by:
\begin{equation}
\omega = \frac{(\alpha \beta h)^2 (N-m+n+1)\zeta}{16[(N-m+n)(1-\rho)+1]\overline{\gamma}_1\overline{\gamma}_2}
\end{equation}
Since the relays are impaired, we can also derive a ceiling in terms of the impairment components that limits the capacity as the impairment becomes more severe. This ceiling is given by \cite[Eq.~(37)]{12}:
\begin{equation}
\overline{C_c} = \log_2\left(1 + \frac{e\nu^2}{2\pi(\mu-\nu^2)}  \right)
\end{equation}
\section{Numerical Results and Discussion}
In this section, the derived analytical expressions are compared with the numerical results using Monte Carlo simulations. Correlated Rayleigh channel coefficients are generated using (1). The atmospheric turbulence $I_a$ is generated using the expression $I_a = I_{aX}\times I_{aY}$, where the two independent random variables $I_{aX}$ and $I_{aY}$ follow the Gamma distribution. In addition, the pointing error is simulated by generating the radial displacement $R$ following the Rayleigh distribution and then by applying Eq.~(19). Since the path loss is deterministic, it can be generated using the relation (18). The table below presents the simulation parameters.
\vspace*{-0.5cm}
\begin{table}[h!]
\renewcommand{\arraystretch}{1.3}
\caption{Simulation Parameters}
\label{tab:example}
\centering
\begin{tabular}{|c|c|}
    \hline
    Parameter  &  Value\\
    \hline
    \hline
    $L$    &    1 km\\
    \hline
    $\lambda$    & 1550 nm   \\
    \hline
    $\gamma_{\text{th}}$    & -20 dB   \\
    \hline
    $F_0$ & -10 m\\
    \hline
    $a$    & 5 cm  \\
    \hline
    $w_0$ & 5 mm\\
    \hline
    $\rho$ & 0.9\\
    \hline
    Modulation & CBFSK\\
    \hline
\end{tabular}
\end{table}
\vspace*{-1cm}
\begin{center}
\includegraphics[width=9cm,height=6.5cm]{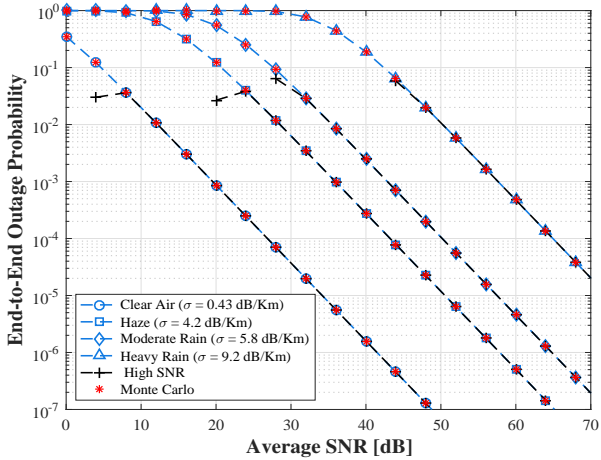}
\captionof{figure}{Outage probability vs. average SNR for different weather conditions and for $C_n^2 = 5 ~10^{-14}, N = m =$ 5, $\sigma_s$ = 3.75 cm, IBO = 30 dB.}
\label{fig1}
\end{center}
\vspace*{-0.5cm}
Fig.~1 shows the variations of the outage probability with respect to the average SNR for various weather conditions. Clearly, we see that the system works better for clear air. However, as the weather changes from a hazy to a rainy situation, the outage performance deteriorates. For an average SNR of 40 dB, the system achieves the following outage values $10^{-6}, 5~10^{-4}, 3~10^{-3}$ and 0.2 for clear air, hazy, moderate and heavy rain, respectively. In fact, the weather attenuation loss comes from the scattering of the signal due to the atmospheric particles. For clear air, the weather is quiet and the scattering loss is negligible or small. Given that the high frequency signals are greatly disturbed by the fog, clouds and dust particles, FSO signal depends not only on the rain which is the major attenuating factor but also on the rate of the rainfall as shown by the figure. In fact, the rain droplets cause a substantial scattering in different directions that mainly attenuate the signal power during the propagation and this phenomena can be explained in more details according to the Rayleigh model of scattering.
\vspace*{-0.7cm}
\begin{center}
\includegraphics[width=9cm,height=6.5cm]{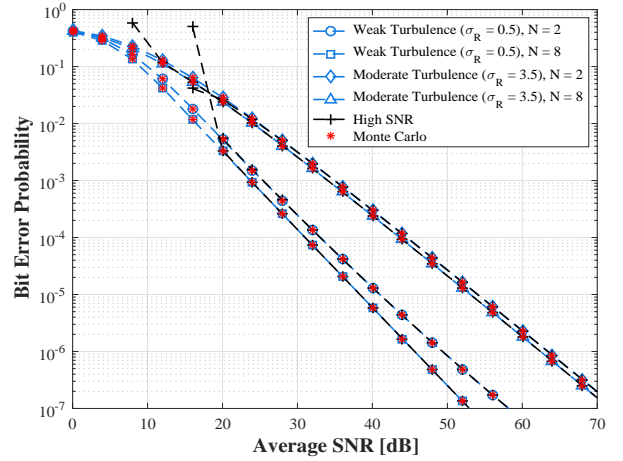}
\captionof{figure}{Bit error probability versus the average SNR for weak and moderate atmospheric turbulences and for ($N = m = 2, N = m = $ 8, clear air ($\sigma$ = 0.43 dB/km), $C_n^2 = 5 ~10^{-14}$, $\sigma_s$ = 3.75 cm, IBO = 30 dB). }
\label{fig1}
\end{center}
\begin{center}
\vspace*{-1cm}
\includegraphics[width=9cm,height=6.5cm]{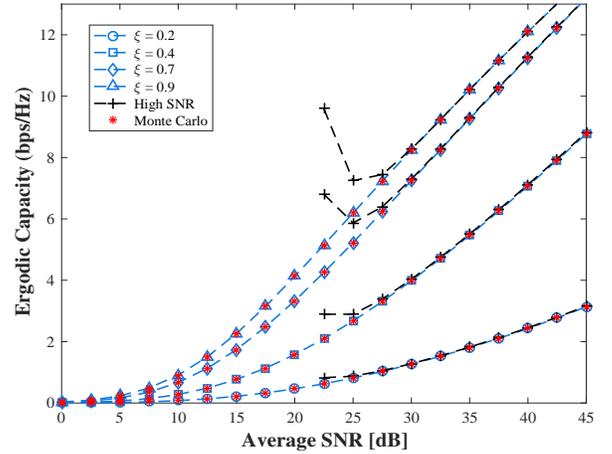}
\captionof{figure}{Ergodic capacity versus the average SNR for different values of the pointing error coefficients and for ($N = m$ = 3, clear air ($\sigma$ = 0.43 dB/km), $C_n^2 = 5 ~10^{-14}$, IBO = 30 dB).}
\label{fig1}
\end{center}
\vspace*{-0.5cm}
Fig.~2 presents the dependence of the average bit error probability on the average SNR under weak and moderate atmospheric turbulence for different number of relays. For weak turbulence, the system performance improves as the number of the relays increases. To achieve an average BEP equal to $10^{-6}$, the system requires the following average SNRs 45, and 49 dB for the number of relays equal to 2 and 8, respectively. In fact, this improvement comes from the spatial diversity (characterized by a large number of relays) that substantially contributes in combating the fading. For moderate turbulence, the system performance completely deteriorates and even increasing the number of the relays, the two performances ($N$ = 2 and 8) are roughly the same. For an average SNR of 50 dB, the system achieves an average BEP approximately equal to $2~10^{-5}$. Thereby, the system depends to a large extent on the state of the optical channel since the number of the relays that characterizes the spatial diversity of the RF channels has no significant impact on the performance under the moderate atmospheric turbulences.\\
Fig.~3 provides the variations of the ergodic capacity against the average SNR for different values of the pointing error coefficients. We observe that the system works better as the pointing error coefficient decreases. In fact, as this coefficient $\xi$ decreases, the pointing error effect becomes more severe. For a given average SNR of 30 dB, the system capacity achieves the following rates 1, 3.9, 7 and 8 bps/Hz for the pointing error coefficients equal to 0.2, 0.4, 0.7 and 0.9, respectively. Thereby, the ergodic capacity gets better as the pointing error coefficient becomes higher.
\vspace*{-0.73cm}
\begin{center}
\includegraphics[width=9cm,height=6.5cm]{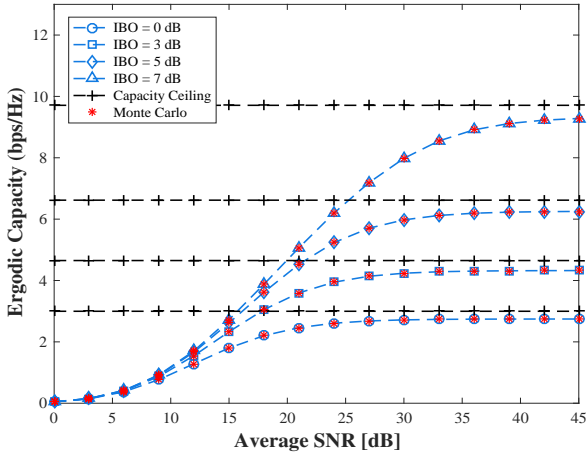}
\captionof{figure}{Ergodic capacity versus the average SNR for different values of IBO and for ($N = m$ = 3, clear air ($\sigma$ = 0.43 dB/km), $C_n^2 = 5 ~10^{-14}$, $\sigma_s$ = 3.75 cm).}
\label{fig1}
\end{center}
\vspace*{-0.5cm}
Fig.~4 shows the variations of the ergodic capacity versus the average SNR for different values of IBO. Clearly, we observe that the ergodic capacity saturates by the ceilings that are caused by the hardware impairments as shown by the figure. In addition, these ceilings disappear for an IBO = 30 dB as shown in Fig.~3 but the performances are limited for the case of lower values of IBO. For the following values of IBO equaling to 0, 3, 5 and 7 dB, the system capacity is saturated by the following ceiling values 3, 4.9, 6.6 and 9.8 bps/Hz, respectively. Note that these ceilings are inversely proportional to the values of the IBO. In fact, as the IBO increases, the saturation amplitude of the relay amplifier increases and thus the distortion effect is reduced. However, as the IBO decreases, i.e, the relay amplifier level becomes lower, the non-linear distortion impact becomes more severe and the channel capacity substantially saturates. Note that the capacity ceiling depends only on the hardware impairment parameters like the clipping factor and the scale of the input signal and not on the system parameters such as the number of the relays and the channels' parameters. Hence, it is straightforward that for any system suffering from the hardware impairments, the channel capacity is always limited by the impairment ceiling regardless of the system configuration such as the channels nature (RF/FSO) and the number of the relays, etc.
\section{Conclusion}
In this work, we investigate a mixed RF/FSO system with multiple relays under the hardware impairments. We derive novel closed-forms of the outage probability, bit error probability, ergodic capacity and high SNR approximation. We evaluate the outage performance in various weather conditions from clear air to more severe state such as heavy rain. We conclude that this performance deteriorates as the weather becomes more severe since the scattering loss increases as the weather worsens. The bit error probability is studied under weak and moderate turbulences for different numbers of relays and it turns out that the system works better especially for large numbers of relays under weak turbulences. Additionally, the system capacity is very sensitive to the pointing error as the coefficient becomes very low. We also quantify the impact of the hardware impairments on the ergodic capacity in terms of the IBO values. We conclude that the capacity saturates more by the impairment ceilings as the IBO decreases. 
\section*{Acknowledgment}
This publication was made possible by NPRP grant 8-408-2-172 from the Qatar National Research Fund (a member of Qatar Foundation). The statements made herein are solely the responsibility of the authors.
\bibliographystyle{IEEEtran}
\bibliography{bibliography}
\end{document}